\documentclass{PoS}
\usepackage{amssymb,amsmath,color}

\title{Radion in Randall-Sundrum model at the LHC and photon collider}

\ShortTitle{Radion in Randall-Sundrum model at the LHC and photon collider}

\author{Yoshiko Ohno
         \\
        Ochanomizu University\\
        E-mail: \email{ohno@hep.phys.ocha.ac.jp}}

\abstract{A warped extra dimension model proposed by Randall and Sundrum
(RS) is one of the attractive candidates to solve the gauge hierarchy
problem in the Standard Model.
In a simplest version of the RS model, there are only two extra
particles beyond the Standard Model - a spin-2 graviton (and its
Kaluza-Klein excitations) and a radion which is a spin-0 and
electrically neutral particle.
In this paper, we discuss a possibility of discovering the radion at a
photon collider, which has been considered as an option of $e^+ e^-$ liner
collider (ILC), focusing on characteristic features of radion interactions
with photon or gluon, in addition to constraints on mass and coupling of
the radion from the SM Higgs boson search at the LHC.
}

\FullConference{KMI International Symposium 2013 on ``Quest for the
                 Origin of Particles and the Universe''\\
                 11-13 December, 2013\\
                 Nagoya University, Japan}

\begin{document}

\section{Introduction}
A warped extra dimension model (so called Randall-Sundrum (RS) model)
has been accepted as an attractive solution of the gauge hierarchy
problem~\cite{Randall:1999ee}. 
This is a model in five-dimensional space time with a warped extra
spatial dimension $y$ which is compactified on the orbifold $S^1/Z_2$. 
There are two 3-branes located at $y=0$ and $\pi r_c$, and they are
called the Planck and the TeV branes, respectively. 
With this set up, the spacetime metric in the RS model is given by 
\begin{eqnarray}
 ds^2=e^{-2k \pi r_c} \eta_{\mu\nu} dx^\mu dx^\nu -dy^2, 
\end{eqnarray}
where $\eta_{\mu\nu}$ is the Minkowski metric and $k$ denotes the
$AdS_5$ curvature. 
Then, the SM Higgs boson mass on the TeV brane could be the weak scale
naturally, i.e., $M_{\rm pl} e^{-k\pi r_c}\sim O(100{\rm GeV})$, when
$kr_c\sim 12$, where $M_{\rm pl}$ is the four dimensional Planck scale. 
%%%------------------
% a new paragrapn

%%%------------------
In the original RS model, only graviton propagates
into the bulk\footnote{
There are variants of the original model which allows (some of) the SM
fields also propagate into the bulk.}. 
Then there are only two types of extra paritlces beyond the SM - a
spin-2 graviton (and its Kaluza-Klein (KK) excitations) and a radion which is
a spin-0 and electrically neutral particle. 
The latter corresponds to fluctuations of a distance between two 
branes. 
When the distance between two branes is not fixed, the radion is
massless. 
However the gauge hierarchy problem requires the distance to be a
certain value, $kr_c\sim 12$. 
The stabilization mechanism of the distance between two branes in the RS
model has been proposed by Goldberger and Wise
(GW)~\cite{Goldberger:1999uk}, and as a result of
the GW mechanism, the radion acquires the mass in the electroweak scale
and it could be a lighter than the 1st KK
graviton~\cite{Goldberger:1999uk, Kribs:2006mq}. 
%%%------------------
% a new paragrapn

%%%------------------
In this presentation, we study production and
decay of the radion in a photon collider, which has been considered as
an option of $e^+ e^-$ linear collider (ILC).
We have shown that the SM Higgs boson search experiments at the LHC give
stringent 
constraints on the radion parameters except for the low-mass region of
the radion~\cite{Cho:2013mva}. 
Then we will show that a photon collider has an excellent potential to 
search for the radion in low-mass region. 

\section{Interactions}
A radion field $\phi$ couples to the trace part of the energy-momentum
(EM) 
tensor of the SM: 
%-----------------
\begin{eqnarray}
 {\cal L}_{\rm int}=\frac{\phi}{\Lambda_\phi}	T^\mu_\mu
\end{eqnarray}
where the scale parameter $\Lambda_\phi$ is expected to be $O({\rm
TeV})$~\cite{Csaki:1999mp, Csaki:2000zn}. 
The trace of EM tensor consists of the SM part
$\left.T^\mu_\mu\right.^{\rm SM}$ and terms from the trace 
anomaly $\left.T^\mu_\mu\right.^{\rm anom}$. The latter is given by 
\begin{eqnarray}
\left.T^\mu_\mu\right.^{\rm anom} = \sum_{a=1,3}\frac{\beta_a(g_a)}{2g_a}F^a_{\mu\nu}F^{a\mu\nu}
\end{eqnarray} 
%-----------------
where $\beta_a(g_a)$ denotes the beta function of QED ($a=1$) and QCD
($a=3$), respectively. 
Note that the interactions of radion with $\left.T^\mu_\mu\right.^{\rm
SM}$ are completely same with the those of the SM Higgs boson replacing 
$\Lambda_\phi$ by the vacuum expectation value $v~(=246~{\rm GeV})$. 
The exception is interactions with $\left.T^\mu_\mu\right.^{\rm
anom}$, which could sizably enhance couplings of the radion with photon
pair and/or gluon pair. 
It is straightforward to calculate production and decay of radion. 
The decay branching ratio of the radion is shown in
ref.~\cite{Cho:2013mva}. 
In general, radion and the SM Higgs boson can mix after the
electroweak symmetry breaking because of scalar-curvature
term from 4-dimensional effective action,
this is called radion-Higgs mixing~\cite{Csaki:2000zn, Giudice:2000av}.
However, we do not consider the mixing because a discovered scalar
particle at the LHC is consistent with the SM~\cite{Aad:2012tfa,
Chatrchyan:2012ufa}.

%-------------------
\section{Production and decay of radion at LHC and photon collider}
As we mentioned earlier, since the radion interactions are similar to
the SM Higgs boson, the mass and scale parameter of
radion, $m_\phi$ and $\Lambda_\phi$, are constrained from the SM Higgs boson
search experiments at the LHC~\cite{Cho:2013mva}. 
Since both ATLAS and CMS reported that there is no signal of any scalar
particles except for the mass $\sim 125~{\rm GeV}$, it could be
read as bounds on the radion parameters. 
We found that the $pp\to h\to ZZ$ process gives the most severe bound on
$(m_\phi, \Lambda_\phi)$. 
For example, $\Lambda_\phi < 8$ TeV is disfavored for $m_\phi \sim 260$ GeV.
It is worth mentioning that the LHC experiments are less sensitive to
search for the radion in low-mass ($\sim 100~{\rm GeV}$) region where 
the decay mode into $\gamma \gamma$ has been measured well rather than
$WW$ and $ZZ$ modes. 
This is because that, due to the trace anomaly, not only the radion
interactions with photons but also with gluons are enhanced
simultaneousely and the decay into $gg$ dominates over the $\gamma
\gamma$ mode. 
%%%------------------
% a new paragrapn

\begin{figure}[htbp]
 \begin{minipage}{0.5\hsize}
  \begin{center}
   \includegraphics[scale=0.2]{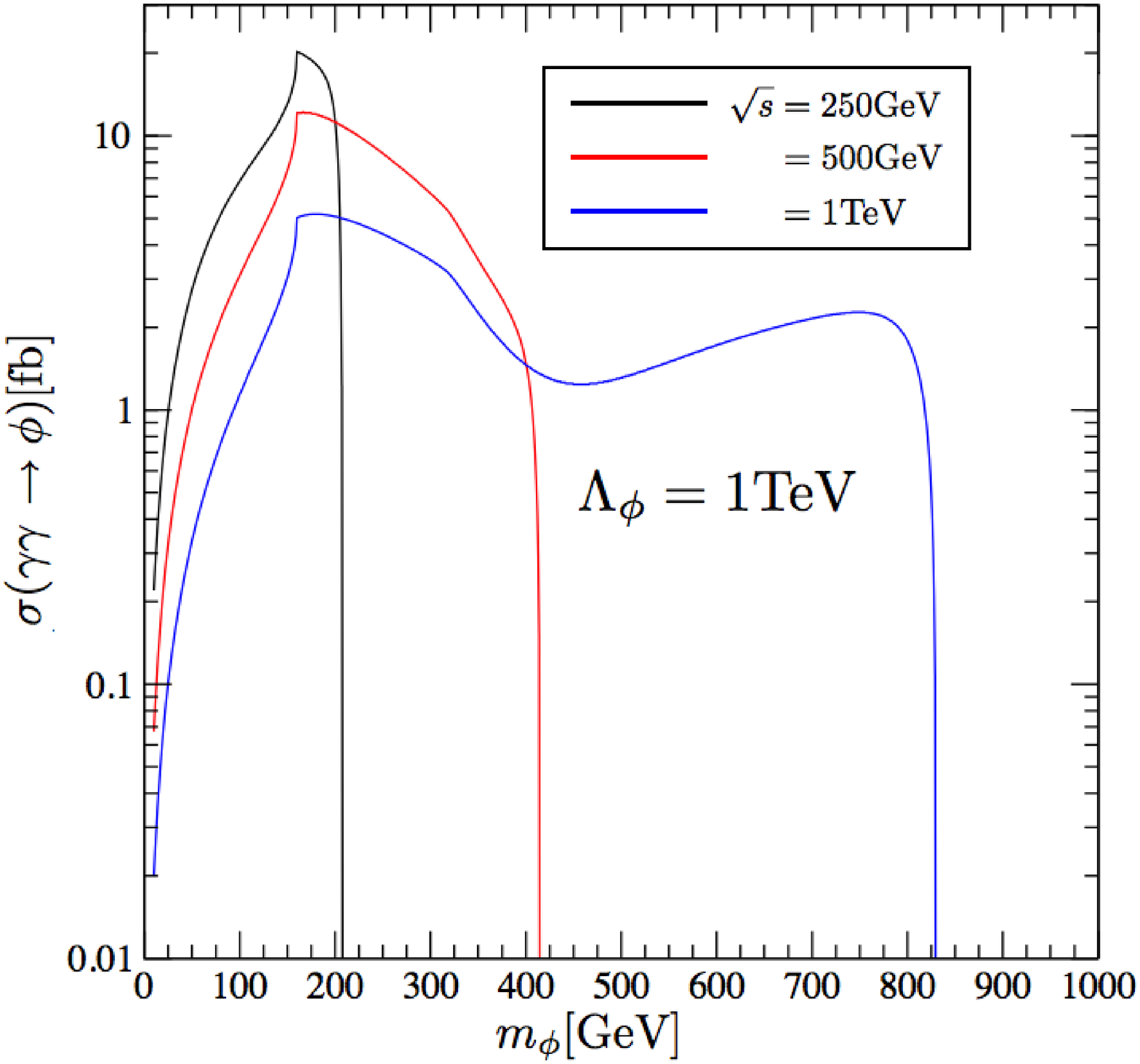}
\quad(a)
  \end{center}
  \label{fig:cs_gam_r_1}
 \end{minipage}
 \begin{minipage}{0.5\hsize}
  \begin{center}
\vspace*{0.1cm}
   \includegraphics[scale=0.2]{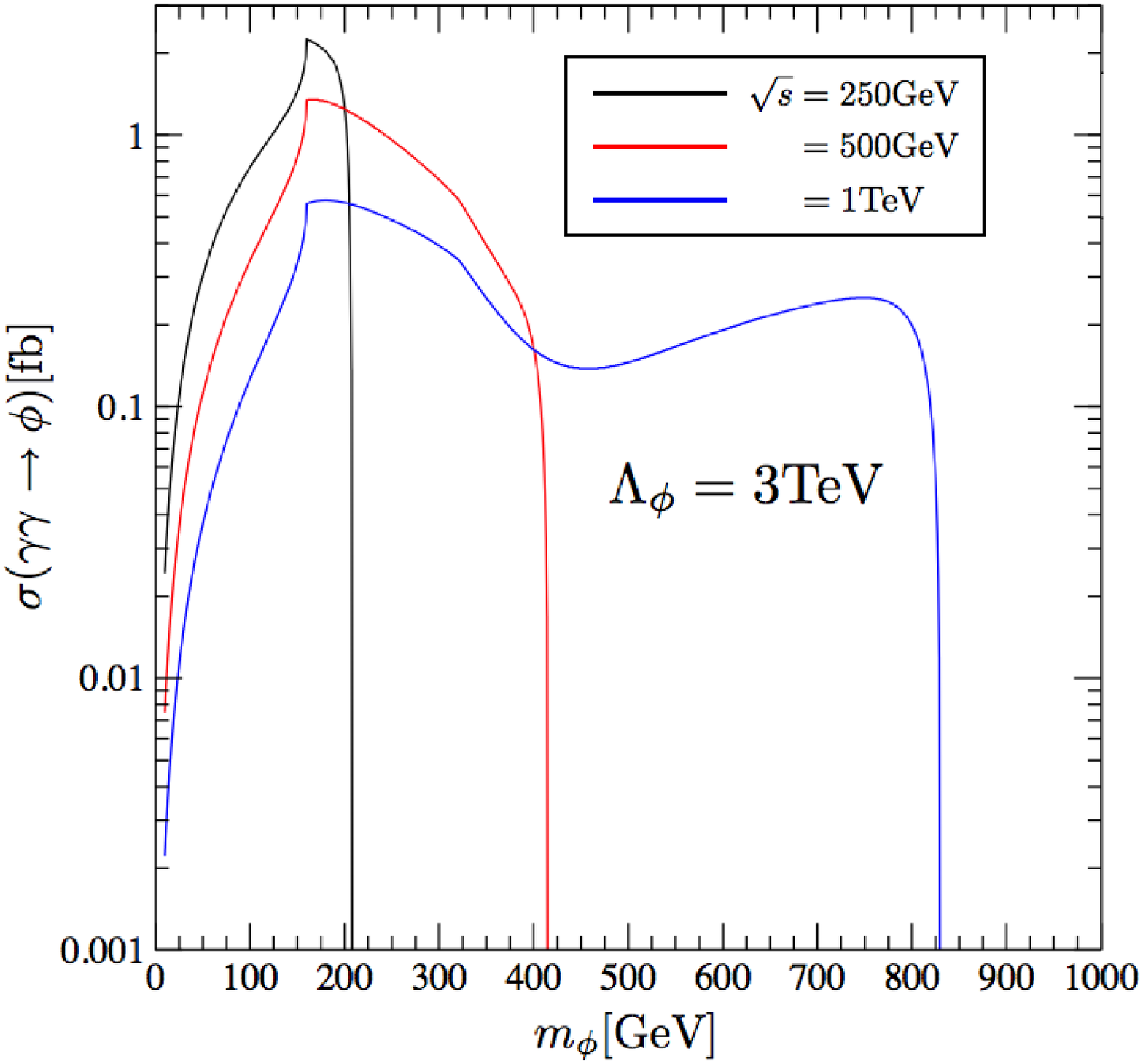}
\quad(b)
  \end{center}
  \label{fig:cs_gam_r_2}
 \end{minipage}
\vspace*{-0.5cm}
%  \caption{The production cross sections of radion at a photon collider
% $e^+e^- \rightarrow \phi$ in the backscattering $\gamma\gamma$ technic
% with center-of-mass $\sqrt{s} = 250{\rm \ GeV}\ ({\rm black}),\ 
% 500{\rm \ GeV}\ ({\rm red})$ and $1{\rm \ TeV}\ ({\rm blue})$. These
% figures (a) and (b) indicate this value for $\Lambda_{\phi}=1, 3$\ TeV,
% respectively.}
  \caption{The production cross sections of radion at a photon collider
 which is convoluted with energy distribution of photon beams
% $e e \rightarrow \phi$ in the backscattering $\gamma\gamma$ technic
 for center-of-mass of electrons $\sqrt{s} = 250{\rm \ GeV}\ ({\rm black}),\ 
 500{\rm \ GeV}\ ({\rm red})$ and $1{\rm \ TeV}\ ({\rm blue})$. These figures (a) and (b) indicate this value for $\Lambda_{\phi}=1, 3$\ TeV,
 respectively.}

  \label{fig:cs}
\end{figure}

%%%------------------
In the photon collider, the radion is produced in a $s$-channel
annihilation of two photons and is dominantly decay into gluon pair when
$m_\phi\sim O(100~{\rm GeV})$. 
We show the production cross section of radion in the photon collider
which is convoluted with energy distribution of photon beams 
in Fig.~\ref{fig:cs} as functions of $m_\phi$ for $\Lambda_\phi=1~{\rm
TeV}$ and $3~{\rm TeV}$. 
Each line corresponds to the collision energy of electron beams 
$\sqrt{s}=250~{\rm GeV}$, $500~{\rm GeV}$ and $1~{\rm TeV}$,
respectively. 
%-----------------
%-----------------
%%%------------------
% a new paragrapn

\begin{figure}[htbp]
 \begin{minipage}{0.5\hsize}
  \begin{center}
   \includegraphics[scale=0.22]{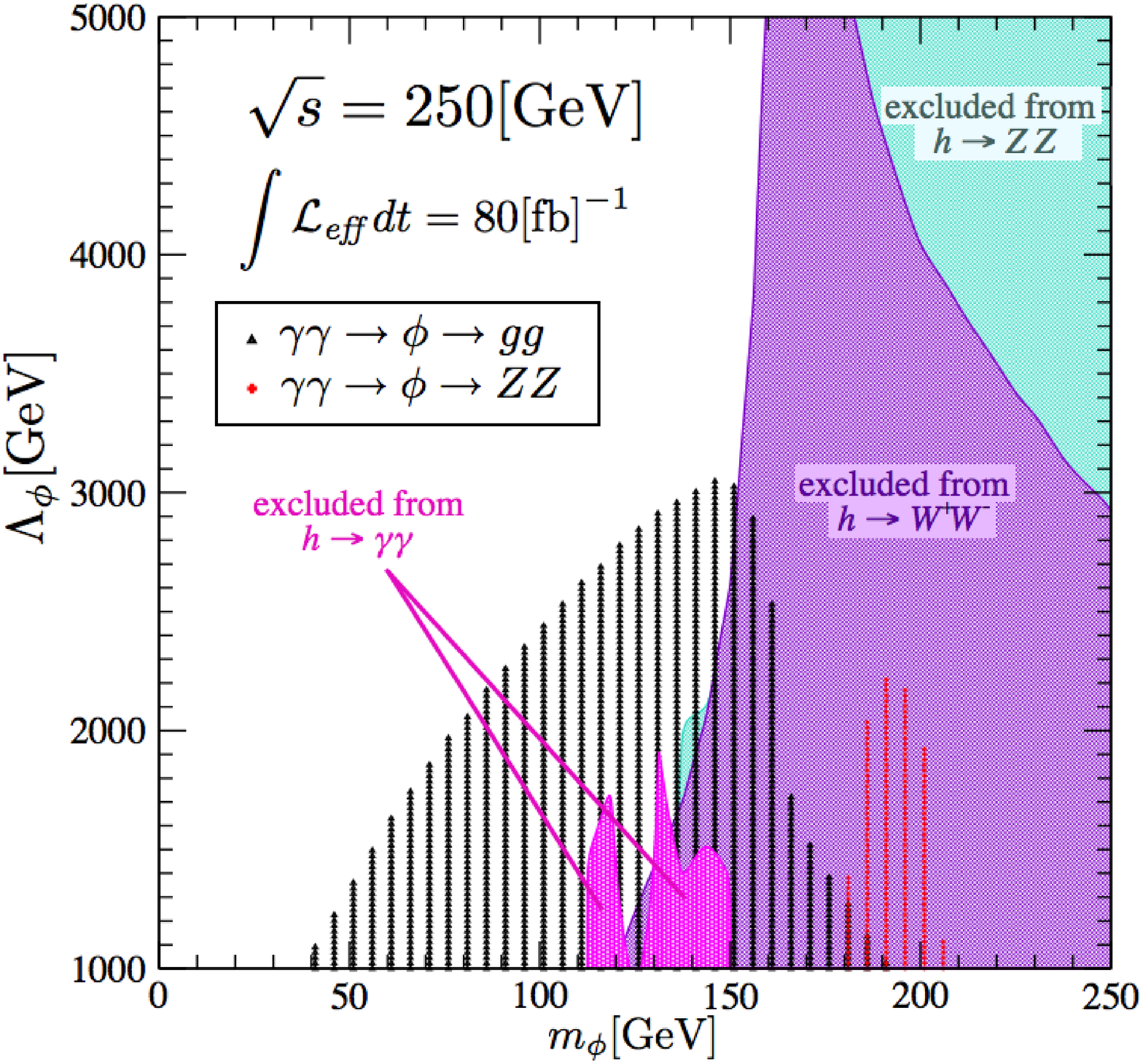}
\quad(a)
  \end{center}
  \label{fig:sig250}
 \end{minipage}
 \begin{minipage}{0.5\hsize}
  \begin{center}
\vspace*{0cm}
   \includegraphics[scale=0.22]{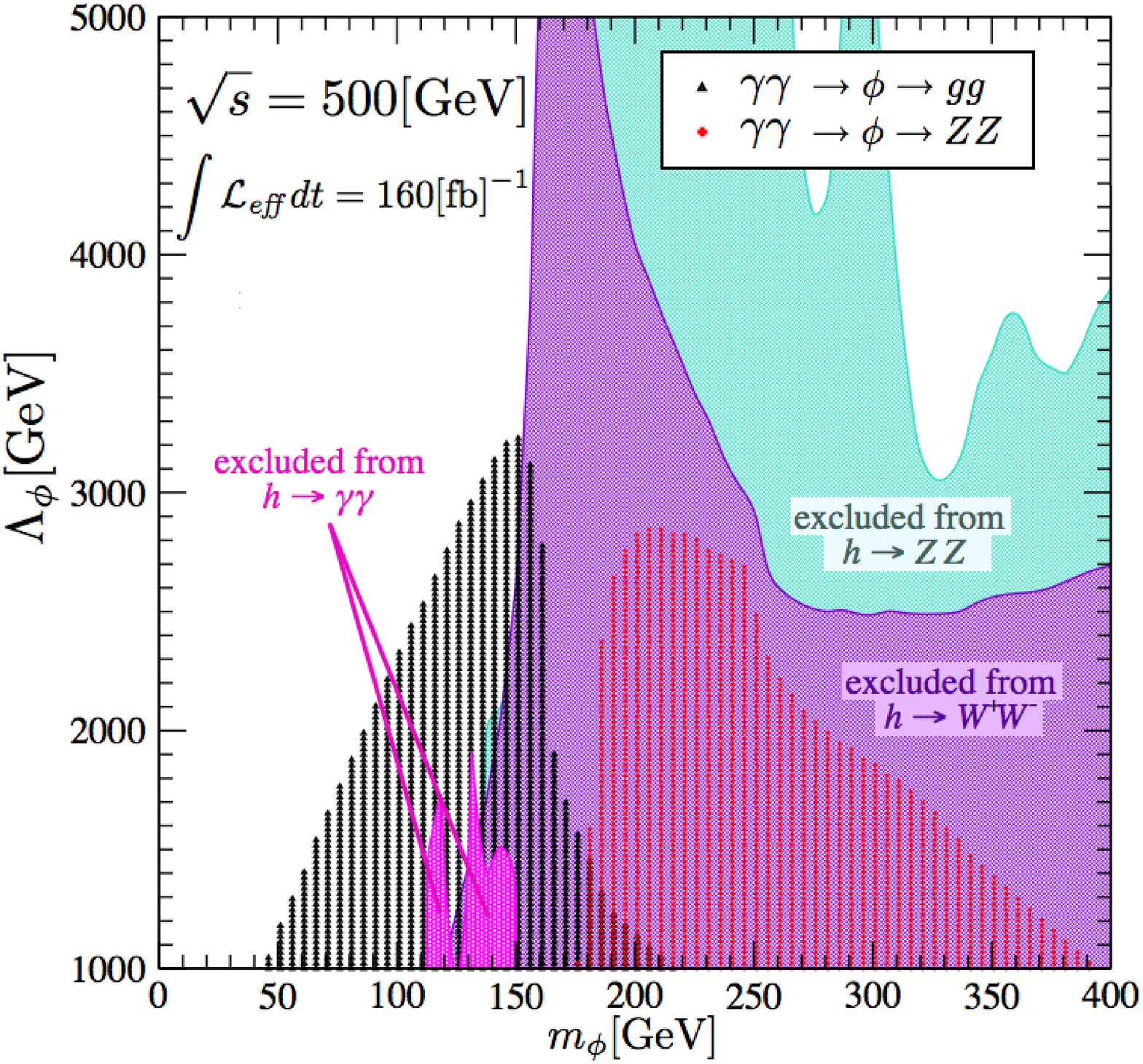}
\quad(b)
  \end{center}
  \label{fig:sig500}
 \end{minipage}
 \begin{minipage}{0.5\hsize}
  \begin{center}
\vspace*{0cm}
   \includegraphics[scale=0.22]{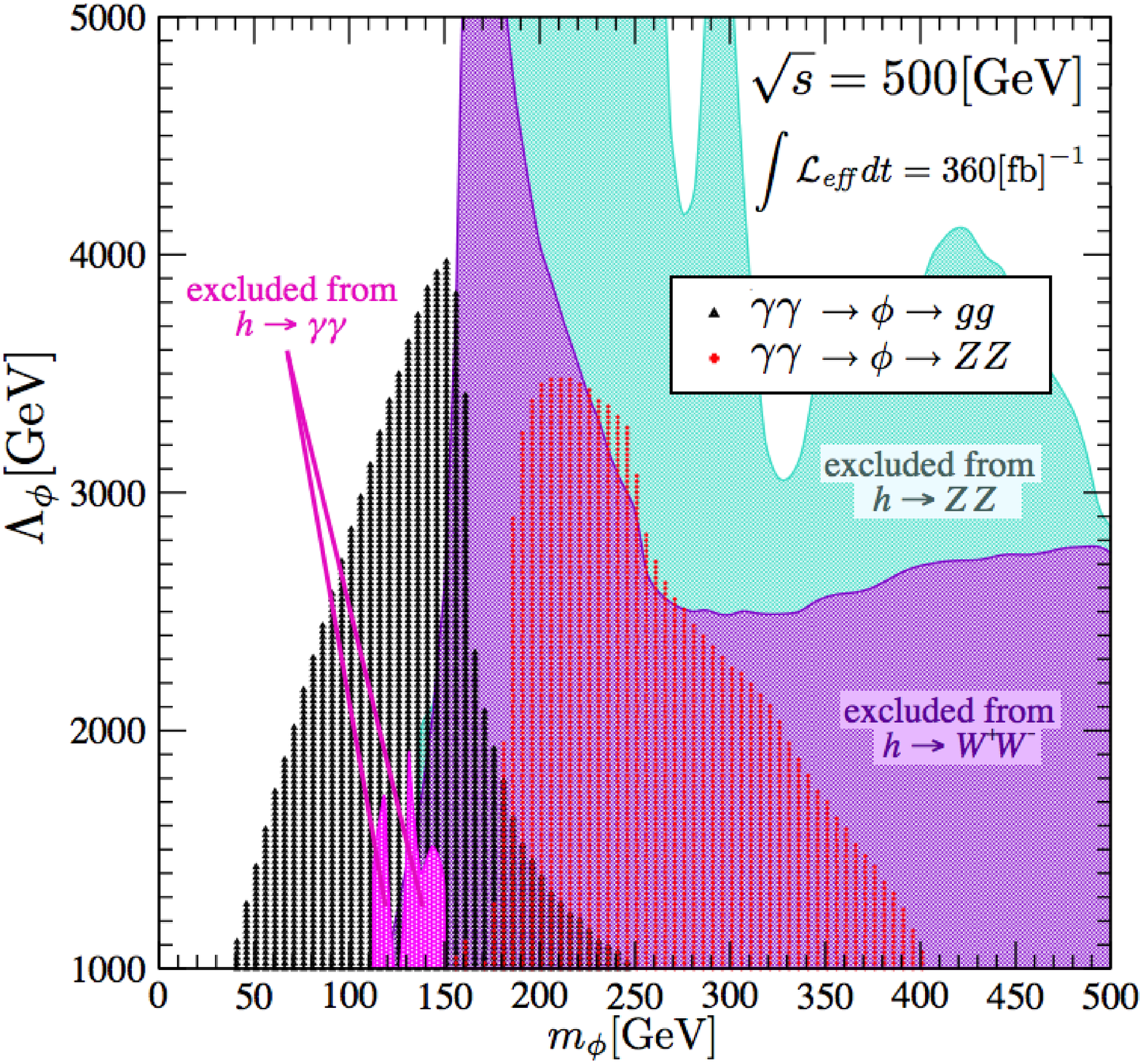}
\quad(c)
  \end{center}
  \label{fig:sig500HL}
 \end{minipage}
 \begin{minipage}{0.5\hsize}
  \begin{center}
\vspace*{0cm}
   \includegraphics[scale=0.22]{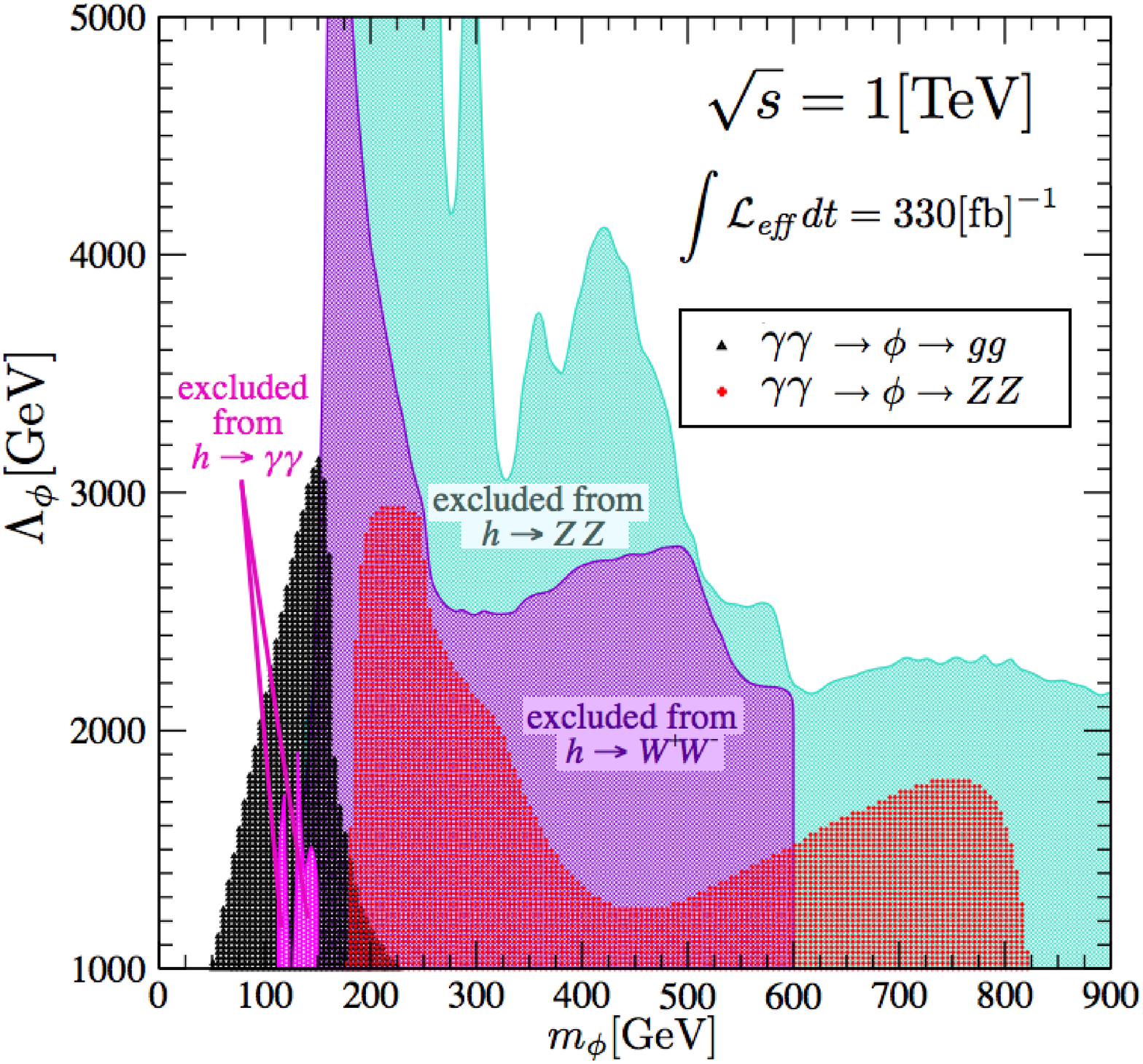}
\quad(d)
  \end{center}
  \label{fig:sig1000}
 \end{minipage}
\vspace*{-0.5cm}
  \caption{
Allowed parameter regions on $(m_\phi,\
 \Lambda_\phi)$ plane at the photon collider for various
 center-of-mass of electrons
 $\sqrt{s}=(a)\ 250$\ GeV,\ (b)\ $500$\ GeV, (c)\ $500$\ GeV\ (with high
 luminosity)\ and (d)\ $1$\ TeV
 which include bounds from the recent result at the
 LHC~\cite{Cho:2013mva}.
 Black and red regions denote parameter regions where
 signal significance is $S/\sqrt{B}>5$ for 
$\gamma \gamma \rightarrow \phi \rightarrow gg$ and
 $\gamma \gamma \rightarrow \phi \rightarrow ZZ$
processes, respectively. Region in turquoise, purple and
 magenta show that the $95\%$ CL excluded region from
 $pp\rightarrow h\rightarrow ZZ,\
 pp\rightarrow h\rightarrow W^+W^-$ and $pp\rightarrow h\rightarrow
 \gamma\gamma$ processes at the LHC.
We assume the effective luminosity $\int \mathcal{L}_{\it eff}\ dt$ is
 $1/3$ of that in ILC~\cite{Behnke:2013xla}, following the TESLA
 technical design report~\cite{Badelek:2001xb}.
}
  \label{fig:sig}
\end{figure}

%%%------------------
Next we examin a possibility of discovering the radion at the photon 
collider assuming the effective integrated luminosity 
$\int \mathcal{L}_{\rm eff} \ dt$. 
We compare the signal process $\gamma \gamma \to \phi \to VV$, 
$(V=g,\gamma,W,Z)$ and the background process $\gamma \gamma \to h \to
VV$, and estimate a significance which is defined by a ratio of the 
signal events $S$ and a square root of background events $\sqrt{B}$. 
The regions on $(m_\phi, \Lambda_\phi)$ plane where the significance
$S/\sqrt{B}>5$ are shown in Fig.~\ref{fig:sig}. 
There are sizable parameter regions where the radion is expected to be
discovered. 
For example, when $m_\phi \sim 150~{\rm GeV}$, the radion can be found
up to $\Lambda_\phi \sim 3~{\rm TeV}$ for $\sqrt{s}=250,\,500\,{\rm
GeV}$ and $1 {\rm TeV}$. 
The excluded region of the radion in low-mass region from the LHC
experiments in $\gamma\gamma$ mode is also shown in the figure. 
We can see that the photon collider is much sensitive than the LHC for
$m_\phi \lesssim 150\,{\rm GeV}$. 

%-----------------
%-----------
\section{Summary}

We studied production and decay of the radion in the
Randall-Sundrum model at the LHC and the photon collider as an option of
$e^+e^-$ linear collider (e.g. the ILC).
%We have shown in Sec.2,
%% contrary to in the case of SM Higgs, 
%the couplings of radion to massless
%gauge bosons (gluon, photon) are enhanced by the trace anomaly.
As a result of our numerical analysis, we found the significance $S/\sqrt{B}>5$
could be expected at the photon collider in sizable parameter region.
For example, when $m_\phi \sim 150~{\rm GeV}$, the radion can be found
up to $\Lambda_\phi \sim 3~{\rm TeV}$ for $\sqrt{s}=250,\,500\,{\rm
GeV}$ and $1 {\rm TeV}$.

\noindent
{\bf\large Acknowledgements}
\vspace{4mm}

The author thanks G.~-C.~Cho and  D.~Nomura for collaborations which
this presentation is based upon.

\end{document}